\documentclass[preprint,aps,12pt,notitlepage,nofootinbib,tightenlines]{revtex4}
\usepackage{amsmath}
\usepackage{bm}
\usepackage{times}
\usepackage{braket}
\usepackage{color}
\usepackage{epsfig}
\usepackage{slashed}
\usepackage{hyperref}
\usepackage{multirow}
\newcommand{\beq}{\begin{eqnarray}}
\newcommand{\eeq}{\end{eqnarray}}

\def \cpc{ {\bf Chin. Phys. C} }
\def \csb{ {\bf Chin. Sci. Bull.} }

\def \ijmpa{ {\bf Int. J. Mod. Phys. A}  }
\def \copc{ {\bf Comput. Phys. Commum. } }
\def \epjc{{\bf Eur. Phys. J. C} }

\def \npb{ {\bf Nucl. Phys. B} }
\def \plb{ {\bf Phys. Lett. B} }

\def \prd{ {\bf Phys. Rev. D} }
\def \prl{ {\bf Phys. Rev. Lett.}  }

\def \jhep{ {\bf JHEP}  }
\def \appb{ {\bf Acta Phys. Polon. B} }
\definecolor{Red}{rgb}{1.,0.,0.}

\definecolor{Blue}{rgb}{0.,0.,1.}

\definecolor{nicered}{rgb}{0.7,0.1,0.1}
\definecolor{nicegreen}{rgb}{0.1,0.5,0.1}
\def\lsim{ {\ \lower-1.2pt\vbox{\hbox{\rlap{$<$}\lower6pt\vbox{\hbox{$\sim$}}}}\ } }
\def\gsim{ {\ \lower-1.2pt\vbox{\hbox{\rlap{$>$}\lower6pt\vbox{\hbox{$\sim$}}}}\ } }

\hypersetup{colorlinks,citecolor=nicegreen,linkcolor=nicered}

\begin{document}
\title{Searches for anomalous $tqZ$ couplings from the trilepton signal of $tZ$ associated production at the 14 TeV LHC}
\author{Jie-Fen Shen$^{1}$, Yu-Qi Li$^{2}$, Yao-Bei Liu$^{2}$\footnote{E-mail: liuyaobei@hist.edu.cn}}
\affiliation{1. School of Biomedical Engineering, Xinxiang Medical University, Xinxiang  453003, P.R.China\\
2. Henan Institute of Science and Technology, Xinxiang 453003, P.R.China }

\begin{abstract}
We investigate the observability of the top anomalous $tqZ$ couplings via the trilepton signatures at the Large Hadron Collider~(LHC) with the center-of-mass energy of 14 TeV. We focus on signals of the $tZ$ associated production with the decay mode $t\to W^{+}b\to b\ell^{+}\nu_{\ell}$, $Z\to \ell^{+}\ell^{-}$, and $t\bar{t}$ production with the decay mode $\bar{t}\to Z(\to \ell^{+}\ell^{-})\bar{q}$ and $t\to b\ell^{+}\nu_{\ell}$, where $\ell=e, \mu$ and $q$ reflects up and charm quarks. It is shown that at $3\sigma$ level, the FCNC top quark decay branching ratios can be probed at, respectively, about $Br(t\to uZ) \leq 1.3\times 10^{-4}$ and $Br(t\to cZ) \leq 4.2\times 10^{-4}$  with the integrated luminosity of 100 fb$^{-1}$, and probed down to $Br(t\to uZ) \leq 2.2\times 10^{-5}$ and $Br(t\to cZ) \leq 8\times 10^{-5}$ for the high-luminosity LHC with 3000 fb$^{-1}$.
\end{abstract}

\maketitle

\newpage
\section{Introduction}
As the most massive particle in the standard model (SM), the top quark is generally considered as an appropriate probe for the new physics (NP) beyond the SM~\cite{top-NP}. In particular, its flavor-changing neutral current (FCNC) interactions are extremely weak in the SM due to the Glashow-Iliopoulos-Maiani (GIM) mechanism~\cite{gim}.
For instance, the branching ratios of $t\to Zu(c)$ are predicted at the order of $10^{-17}(10^{-14})$ in the SM~\cite{13112028}.
However,  several extensions of the SM such as the SUSY models~\cite{susy1,susy2}, two-Higgs-doublet models~\cite{2hdm}, extra dimensions~\cite{ED}, and the other miscellaneous models~\cite{other} predict
much higher branching ratios up to $10^{8}-10^{10}$ order of magnitude larger than SM predictions. Therefore, any signal for these rare FCNC processes at a measurable rate would be a robust evidence for NP beyond the SM.

Over the years, the top quark FCNC interactions has been studied intensively via the $t\bar{t}$ production processes with the anomalous decays of top quarks or anomalous production of single top quark~\cite{t1,t2,t3}. Furthermore, the anomalous top quark interactions affect $b$ quark FCNC decays through loop diagrams as mentioned in Ref.~\cite{b-decay}.
Very recently, both the ATLAS and the CMS experiments have obtained the limits on the branching ratios of the top anomalous decays through different channels (for an updated review, see \cite{moriond-2017}). The current upper limits for $Br(t\to Zq)$ at $95\%$ confidence level (CL) have been found to be~\cite{cms1,atlas1}:
\beq
Br(t\to Zu)\leq\left\{ \begin{array}{ll}
2.2\times 10^{-4} & {\rm CMS}, \\
7\times 10^{-4} & {\rm ATLAS}, \\ \end{array} \right. \label{eq:tuz} \quad
Br(t\to Zc)\leq\left\{ \begin{array}{ll}
4.9\times 10^{-4} & {\rm CMS}, \\
7\times 10^{-4} & {\rm ATLAS}. \\ \end{array} \right. \label{eq:tcz}
\eeq
The most stringent bounds on the strengths of anomalous couplings $tqZ$
come from the CMS experiment with $\sqrt{s}=8$ TeV, using the recent combination with anomalous $tZ$ production~\cite{cms1}. It is notable to mention here that, even at the future facilities, these bounds resulting from $t\bar{t}$ production would not be improved considerably. For example, the upper bounds on the branching ratios of $t\to u(c)Z$ by the ATLAS Collaboration
are $1.3~(2.3)\times 10^{-4}$ at $95\%$ CL for the high-luminosity Large Hadron Collider (HL-LHC) with 3000 fb$^{-1}$ at 14 TeV~\cite{atlas-14-3000}. Therefore, we here concentrate on the associated production of a $Z$ boson with a single
top quark at the 14 TeV LHC.

The aim of this letter is to investigate the limits on anomalous $tZq$ couplings  by considering $tZ$ associated production.
Compared with the pioneering study focusing on 14 TeV LHC~\cite{t2}, we make use of a more accurate description of the signal and SM backgrounds relying on the advanced Monte Carlo event generation and the detector simulation. We investigate the final state of trilepton, where the top quark decays to a charged lepton, a $b$ quark and neutrino and the $Z$ boson decays into a pair of leptons.  On the other hand, we also consider the $t\bar{t}$ production process with the top quark semilepton decay and the antitop decaying into $Z\bar{q}$ via the anomalous $tZq$ vertex. It has the same signal if the light quark is missed by the detector.

The organization of this paper is as follows. In Sec.~II, we present the theoretical framework which
describes the FCNC $tZq$ couplings. In Sec.~III, we discuss the signals of $tZ$ associated production with the decay mode $t\to W^{+}b\to \ell^{+}\nu b$ and $Z\to \ell^{+}\ell^{-}$, and $t\bar{t}$ production with the decay mode $\bar{t}\to Z(\to \ell^{+}\ell^{-})\bar{q}$ and $t\to b\ell^{+}\nu_{\ell}$. Then we analyze the sensitivity of 14 TeV LHC to anomalous $tqZ$ couplings in detail. Finally, we conclude in Sec.~IV.

\section{Calculation framework}
In general, the effective Lagrangian describing the interactions between the top quark and a light up-type quark ($u$ or $c$) and the $Z$ boson can be written as~\cite{08113842}
\begin{eqnarray}
\label{lag}
-\mathcal{L}_{eff} &=&\sum_{q=u,c} [\frac{g}{2c_{w}}\kappa_{tqZ} \bar{q}\frac{i\sigma^{\mu\nu}q_{\nu}}{\Lambda} (\kappa_{L} P_{L}+\kappa_{R} P_{R})tZ_{\mu} \nonumber\\
&+& \frac{g}{2c_{w}}\lambda_{tqZ} \bar{q}\gamma^{\mu} (\lambda_{L} P_{L}+\lambda_{R} P_{R})tZ_{\mu}] + H.c.,
\end{eqnarray}
where $g$ is the $SU(2)_L$ gauge coupling constant, $C_{W}=\cos\theta_{W}$ and $\theta_{W}$ is the Weinberg
angle, $P_{L,R}=\frac{1}{2} (1\mp \gamma^5)$, $\sigma^{\mu\nu} = \frac{1}{2}[\gamma^{\mu},\gamma^{\nu}]$, and $\Lambda$ is the new physics scale, which is related to the cutoff mass scale above which the effective theory breaks down. The effects of new physics contributions are quantified through the dimensionless parameters $\kappa_{tqZ}$ and $\lambda_{tqZ}$ together with the complex chiral parameters $\kappa_{L,R}$ and $\lambda_{L,R}$,
 which are normalized as $|\kappa_{L}|^2 + |\kappa_{R}|^2=|\lambda_{L}|^2 + |\lambda_{R}|^2=1$.

 The above effective Lagrangian can be used to calculate both production cross sections and the branching ratios of the $t\to qZ$ decays.
 Note that we do not consider the FCNC $tqg$ couplings because the sensitivity is poor in comparison to other channels~\cite{atlas2}. On the other hand,
the $\lambda_{tqZ}$ couplings lead to very small cross sections~\cite{plb725-123}. We thus only consider the cases where $\kappa_{tqZ}/\Lambda\neq 0$, and no specific chirality is assumed for the FCNC interaction vertices, i.e. $\kappa_{L}=\kappa_{R}$.

 At the leading order (LO) and the next-to-leading order~(NLO), the decay widths of the dominant top quark decay mode $t\rightarrow Wb$ could be found in Ref.~\cite{twb}.
 The partial decay widths of $t\to qZ$  with flavor-violating interactions are given by
\begin{eqnarray}
\Gamma (t \rightarrow qZ) = \frac{\alpha}{32 s_W^2 c_W^2} m_t^3 \frac{|\kappa_{tqZ}|^2}{\Lambda^2}
\left[1-\frac{m_{Z}^2}{m_{t}^2}\right]^2 \left[2+\frac{m_{Z}^2}{m_{t}^2}\right]
\end{eqnarray}
 After neglecting all the light quark masses and assuming the dominant top decay width $t \to bW$, the branching ratio of $t \to qZ$ can be approximately given by:
 \beq
Br(t \to qZ)=5.74\times 10^{-3}|\kappa_{tqZ}(\rm 1 TeV)/\Lambda|^2.
\eeq
Here the NLO QCD correction to the top quark decay via model-independent FCNC couplings is also included and the $k$-factor is taken as 1.02~\cite{top-nlo}. The SM input parameters relevant in our study are taken as follows~\cite{pdg}:
\begin{align}
m_t=173.1{\rm ~GeV}, \quad m_Z=91.1876{\rm ~GeV}, \quad s_W^2=0.223, \quad \alpha&=1/127.94.
\end{align}

\section{Signal and discovery potentiality}
\begin{figure}[htb]\vspace{0.5cm}
\begin{center}
\centerline{\epsfxsize=16cm \epsffile{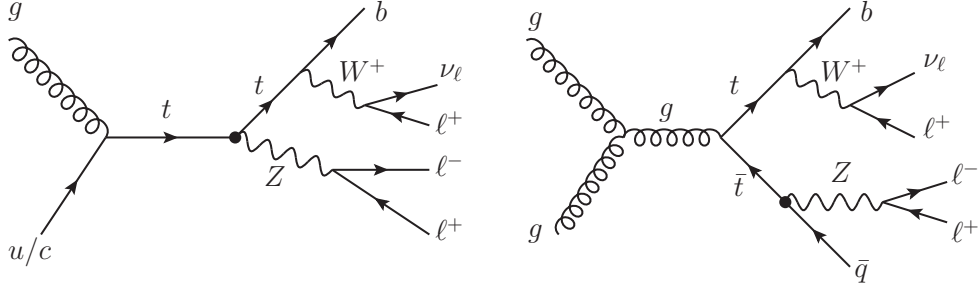}}
\vspace{-16cm}
\caption{Representative Feynman diagrams for $tZ$ production and $t\bar{t}\to tZ\bar{q}$ production via FCNC $tZq$ couplings.}
\label{fey}
\end{center}
\end{figure}

In this section, we perform the Monte Carlo simulation and explore the sensitivity of 14 TeV LHC to the $tqZ$ FCNC couplings through the $tZ$-FCNC and $t\bar{t}$-FCNC processes. The representative Feynman diagrams for the signal processes are shown in Fig.~\ref{fey}.

Obviously, the signal is taken as the trilepton plus one $b$-jet and missing energy. The main backgrounds which yield the identical final states to the signal are $t\bar{t}$, $t\bar{t}V$ ($V=W, Z$), $WZ+$ jets and the irreducible $tZj$, where $j$ denotes non-bottom-quark jets. In the $t\bar{t}$ case (both top quarks decay semi-leptonically), a third lepton comes from a semi-leptonic $B$-hadron decay in the $b$-jet. Here we do not consider multijet backgrounds where jets can be faked as electrons, since they are very negligible in multilepton analyses~\cite{14067830}.  On the other hand, the SM $t\bar{t}h$ and tri-boson events can also be the sources of backgrounds for our signal. We have not included these backgrounds in the analysis due to very small cross sections after applying the cuts.
 The high order corrections for the dominant backgrounds are considered by including a $k$-factor, which is 2.07 for $WZ+$ jets~\cite{1604.08576}, 1.27 for $t\bar{t}V$~\cite{nlo-ttv} and 1.7 for $tZj$~\cite{nlo-tzj}, respectively. The LO $t\bar{t}$ samples are normalized to the theoretical cross-section
    value for the inclusive $t\bar{t}$ process of 953.6 pb performed at
    next-to-next-to-leading order (NNLO) in QCD and including resummation
    of next-to-next-to-leading logarithmic (NNLL) soft gluon terms~\cite{1303.6254}.  On the other hand, the MLM matching scheme is used, where we included up to three extra jets
    for $WZ$ + jets and $tZj$ in the simulations~\cite{MLM}. Here it should be mentioned that the $k$-factor for the LO cross section of $\sigma_{tZ}$ is chosen as about 1.4 at the 14 TeV LHC~\cite{prd83-114049,nlo-tz}.

In order to simulate and generate the signal events, the $\kappa$ Lagrangian terms presented in Eq.(\ref{lag}) are implemented in MadGraph5-aMC$@$NLO~\cite{mg5} by means of the FeynRules package~\cite{feynrules}.
All of these signal and backgrounds events are generated at LO with the CTEQ6L parton distribution function (PDF)~\cite{cteq}, and the renormalization and factorization scales are set dynamically by default. The events are then passed to Pythia 6~\cite{pythia} for parton showering and hadronization, and the fast detector simulation in Delphes~\cite{delphes} with CMS detector card is used to include the detector effects.
Finally, events are  analyzed by using the program of MadAnalysis5 \cite{ma5}.

Firstly, we employ some basic cuts on the signal and background events:
\begin{itemize}
\item Basic cuts: $p_{T}^{\ell} > 20 \rm ~GeV$, $p_{T}^{j, b} > 25 \rm ~GeV$, $|\eta_{\ell, j, b}|<2.5$,  where $\ell=e, \mu$.
\end{itemize}
 Further, we apply some general preselections as follows.
 \begin{itemize}
\item Cut-1: There are exactly three isolated leptons ($N(\ell)=3$) and exactly one b-tagged jet ($N(b)= 1$).
\end{itemize}
 The requirement of three leptons can strongly reduce the $t\bar{t}$ backgrounds, and the $b$-tagging can efficiently suppress the diboson components.

\begin{figure}[htb]
\begin{center}
\vspace{0.5cm}
\centerline{\epsfxsize=14cm\epsffile{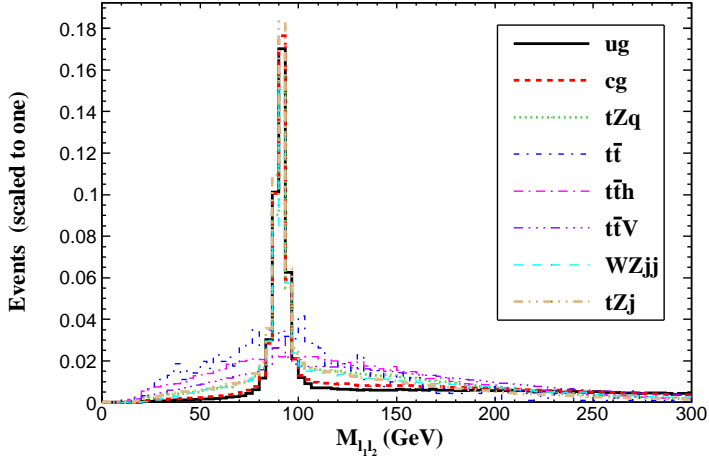}}
\caption{Normalized invariant mass distribution of the first and second leptons at 14 TeV LHC.}
\label{mz}
\end{center}
\end{figure}

In Fig.~\ref{mz}, we show the transverse momentum distributions of two leptons, labeled by $\ell_{1}$ and $\ell_{2}$, in the signal and backgrounds at 14 TeV LHC. Two of the same-flavour leptons in each event are required to have
opposite electric charge, and have an invariant mass, $M_{\ell_{1}\ell_{2}}$, compatible with the $Z$ boson mass, i.e.,
 \begin{itemize}
\item Cut-2: $|M(\ell_{1}\ell_{2})-m_{Z}|< 15 \rm ~GeV$.
\end{itemize}

\begin{figure}[htb]
\begin{center}
\vspace{0.5cm}
\centerline{\epsfxsize=14cm \epsffile{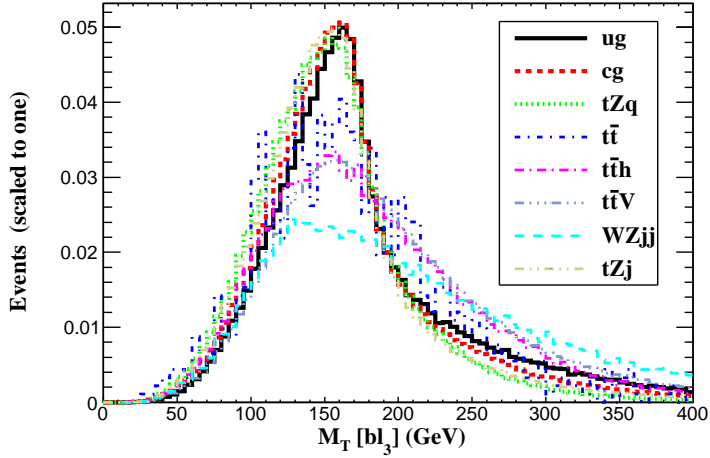}}
\caption{Normalized transverse mass distribution for the $b\ell_{3} \slashed E_T$  system at 14 TeV LHC.}
\label{mt}
\end{center}
\end{figure}

Since the third lepton, $\ell_{3}$, is assumed to originate from the leptonically decaying top quark, the top quark transverse cluster mass could be defined as~\cite{epjc-74-3103}
\beq
M_T^{2}\equiv(\sqrt{(p_{\ell_{3}}+p_{b})^{2}+|\vec{p}_{T,\ell_{3}}+\vec{p}_{T,b}|^{2}}+|\vec{\slashed p}_T| )^{2}-|\vec{p}_{T,\ell_{3}}+\vec{p}_{T,b}+\vec{\slashed p}_T|^{2},
\eeq
where $\vec{p}_{T,\ell_{3}}$ and  $\vec{p}_{T,b}$ are the transverse momentums of the third charged leptons and $b$-quark, respectively, and $\vec{\slashed p}_T$ is the missing transverse momentum determined by the negative sum of visible momenta in the transverse direction.
In Fig.~\ref{mt}, we show the transverse mass distribution for the $b\ell_{3} \slashed E_T$ system.
From this figure, we can see that the distributions of signal and backgrounds including top quark have peaks
around the top quark mass. Therefore, we choose the transverse mass $M_T$ cuts
\begin{itemize}
\item Cut-3: $140 \rm ~GeV <M_{T}(b\ell_{3})< 190 \rm ~GeV$.
\end{itemize}

\begin{figure}[htb]
\begin{center}
\vspace{0.5cm}
\centerline{\epsfxsize=13cm \epsffile{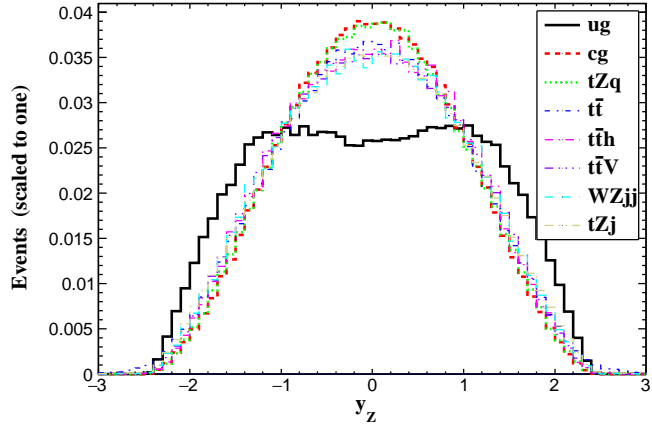}}
\caption{Normalized distribution of the rapidity of the $Z$ boson for the signals and backgrounds.}
\label{yz}
\end{center}
\end{figure}

In Fig.~\ref{yz}, we present the normalized spectrum of the rapidity of the
reconstructed resonances for the signal and backgrounds. It can be seen the $Z$ boson from the $ug\to tZ$ process concentrates in the forwards and backwards regions. This is because the momentum of initial up quark is generally larger than that of gluon, the partonic center-of-mass frame is highly boosted along the direction of the up quark. This case is similar with the top-Higgs associated production process~\cite{prd86-094014}. Thus we impose rapidity cut on reconstructed $Z$ boson for the signal of $ug\to tZ$ process as
\begin{itemize}
\item Cut-4: $|y_{Z}| > 1.0$.
\end{itemize}

The cross sections of the signal and backgrounds after imposing
the cuts are summarized in Table~\ref{cutflow}, the anomalous couplings are chosen to be
$\kappa_{uZ}(\rm 1 TeV)/\Lambda=\kappa_{cZ}(\rm 1 TeV)/\Lambda=0.1$. One can see that all the backgrounds are suppressed very efficiently after imposing the selections.
However, the cross section of the process $pp\to t\bar{t}\to tZ\bar{c}$ is about two times larger than that of $cg\to tZ$ process after cuts. As stated before, we should include these two processes when discussing the $tcZ$ couplings.
On the other hand, since the momentum of initial charm quark is much smaller than that of the initial up quark, the $Z$ boson from $cg$ initial states is
not boosted as from $ug$ initial states.  Therefore, we do not apply the cut-4 when it comes to the $cg\to tZ$ process.

\begin{table}[htb]
\begin{center}
\caption{The cut flow of the cross sections (in fb) for the signal and backgrounds at the 14 TeV LHC. The anomalous couplings are chosen to be
$\kappa_{uZ}(\rm 1 TeV)/\Lambda=\kappa_{cZ}(\rm 1 TeV)/\Lambda=0.1$. \label{cutflow}}
\vspace{0.2cm}
\begin{tabular}{|c|c|c|c|c|c|c|c|}
\hline
Cuts & $ug\to tZ$ & $cg\to tZ$ & $t\bar{t}\to tZq$ & $t\bar{t}$  & $t\bar{t}V$  & $WZjj$ & $tZj$ \\ \hline
Basic cuts & 1.42 & 0.15 & 0.61& 11851  & 1.63 & 7.24 & 3.71 \\ \hline
Cut-1 & 0.45 & 0.049 & 0.083 & 0.69 & 0.16 & 1.88 & 0.59 \\ \hline
Cut-2 &0.44&0.047&0.081&0.37&0.077&1.85&0.58\\ \hline
Cut-3& 0.27&0.027&0.041&0.14&0.024&0.46&0.31\\
\hline
Cut-4& 0.14&0.01&0.014&0.04&0.008&0.15&0.11\\
\hline
\end{tabular} \end{center}\end{table}

The statistical significance is calculated after final cut by using~\cite{ss}:
\beq
SS=\sqrt{2\pounds_{int}[(\sigma_S+\sigma_B)\ln(1+\sigma_S/\sigma_B)-\sigma_S]},
\eeq
where $\sigma_S$ and $\sigma_B$ are the signal and background cross sections
 and $\pounds_{int}$ is the integrated luminosity.
Here we define the discovery significance as $SS=5$, the possible evidence as $SS=3$ and the exclusion limits as $SS=2$.
\begin{figure}[htb]
\begin{center}
\vspace{-0.5cm}
\centerline{\epsfxsize=9cm \epsffile{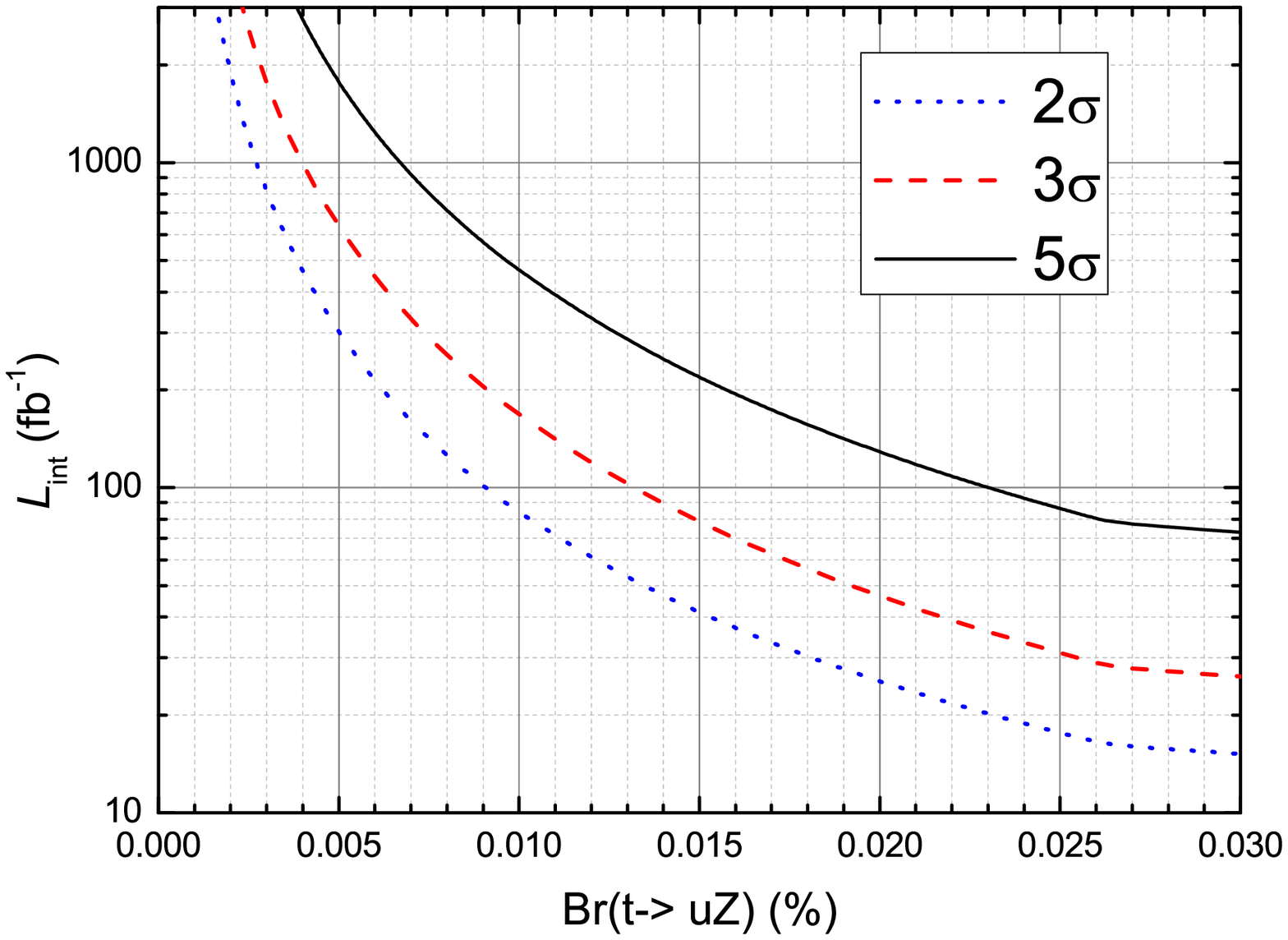}\epsfxsize=9cm \epsffile{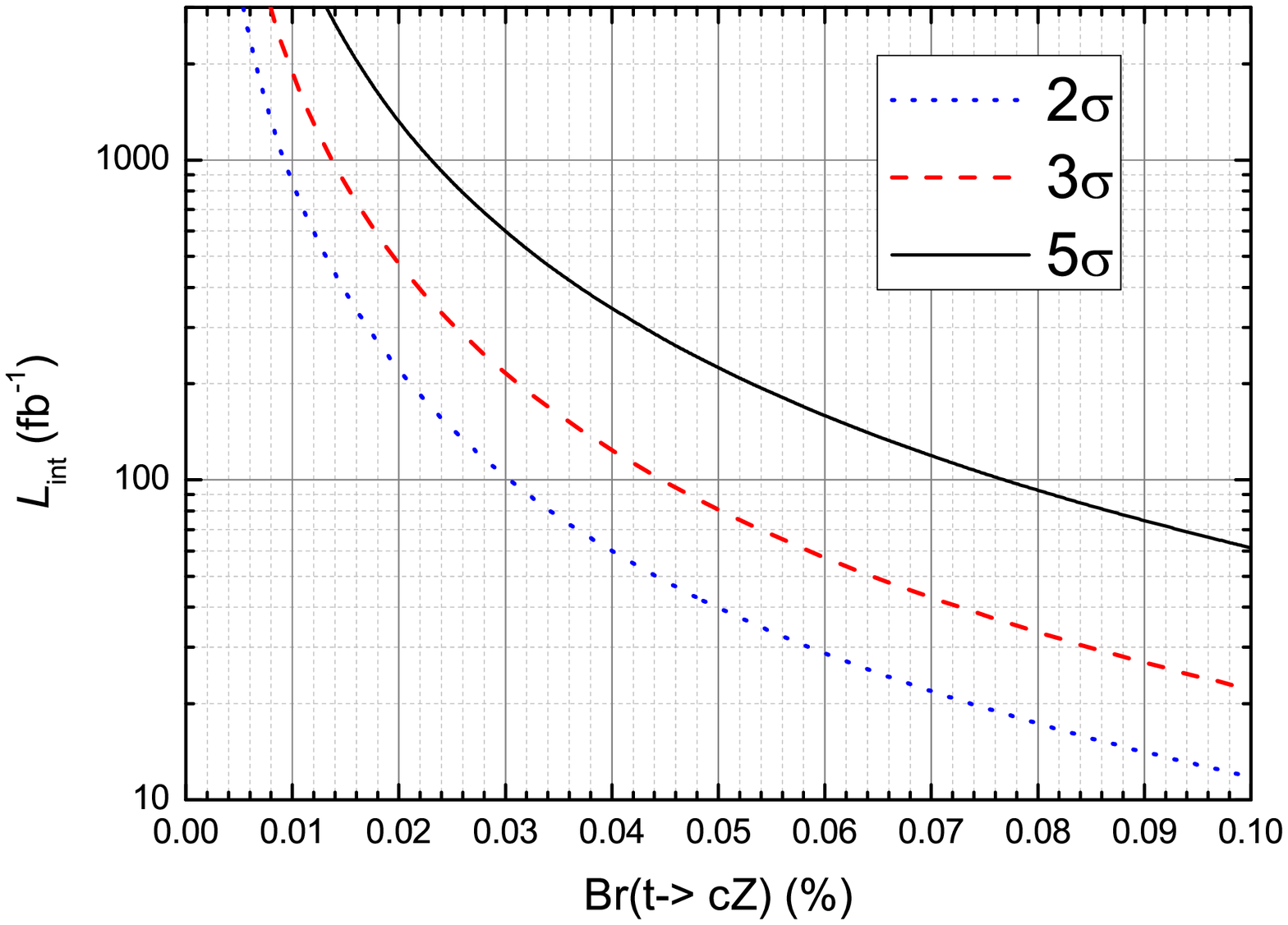}}
\caption{The $2\sigma$ (dotted curves), $3\sigma$ (dashed curves) and $5\sigma$ (solid curves) contour plots for the signal in $L_{int}-Br(t\to uZ)$ (left) and $L_{int}-Br(t\to cZ)$ (right) planes at 14 TeV LHC. }
\label{ss}
\end{center}
\end{figure}

In Fig.~\ref{ss}, the $2\sigma$, $3\sigma$ and $5\sigma$ lines are drawn as a function of the integrated luminosity and the branching ratios $t\to qZ$. We do not consider the theoretical and systematic uncertainties for simplicity. One can see that the $5\sigma$ CL discovery sensitivity of $Br(t\to uZ)$ is about $2.3\times 10^{-4}$ when the integrated luminosity is 100 fb$^{-1}$.
If no signal is observed, it means that the FCNC $tqZ$ couplings can not be too large.
If we take the integrated luminosity as 100 fb$^{-1}$, the $3\sigma$ CL upper limits on the branching ratios $Br(t\to qZ)$ are about $Br(t\to uZ) \leq 1.3\times 10^{-4}$ and $Br(t\to cZ) \leq 4.2\times 10^{-4}$. For the future HL-HLC, the $3\sigma$ CL upper limits on the branching ratios $Br(t\to qZ)$ probed down to $Br(t\to uZ) \leq 2.2\times 10^{-5}$ and $Br(t\to cZ) \leq 8\times 10^{-5}$.
It is remarkable that even with the high-luminosity of 3000 fb$^{-1}$, the branching ratios would not be measured better than $10^{-5}$. The recent phenomenological studies in Ref.~\cite{1709.03975} have shown that the $95\%$ CL upper limits on the branching ratios $Br(t\to qZ)$ probed down to $Br(t\to uZ) \leq 4.1\times 10^{-5}$ and $Br(t\to cZ) \leq 1.6\times 10^{-3}$. Thus our results are comparable with those for the HL-LHC, but they are below the sensitivity limits of the future 100 TeV $pp$ circular collider (FCC-hh)~\cite{fcc}.

\section{CONCLUSION}
In this letter, we have investigated the signal of the $tZ$ associated production via the FCNC $tqZ$ couplings at the LHC with $\sqrt{s}=14$ TeV. We focus on trilepton final signals of the $pp\to tZ$ process with the decay mode $t\to W^{+}b\to b\ell^{+}\nu_{\ell}$, $Z\to \ell^{+}\ell^{-}$, and $t\bar{t}$ production process with the decay mode $\bar{t}\to Z(\to \ell^{+}\ell^{-})\bar{q}$ and $t\to b\ell^{+}\nu_{\ell}$, where $\ell=e, \mu$ and $q$ reflects up and charm quarks. It is shown that the branching ratios $Br(t\to uZ)$ and $Br(t\to cZ)$ are, respectively, about $Br(t\to uZ) \leq 1.3\times 10^{-4}$ and $Br(t\to cZ) \leq 4.2\times 10^{-4}$ at $3\sigma$ level with the integrated luminosity 100 fb$^{-1}$, and probed down to $Br(t\to uZ) \leq 2.2\times 10^{-5}$ and $Br(t\to cZ) \leq 8\times 10^{-5}$ for the future HL-LHC,  which are significantly better than the current experimental results.

\begin{acknowledgments}
This work is supported by the Foundation of He¡¯nan Educational Committee (Grant no. 2015GGJS-059) and the Foundation of Henan Institute of Science and Technology (Grant no. 2016ZD01).
\end{acknowledgments}

\end{document}